\documentclass[12pt]{article}
\usepackage{fullpage}
\usepackage{amsmath}
\usepackage{amssymb}
\usepackage[dvips]{epsfig}

\def\##1{\underline{#1}}
\def\=#1{\underline{\underline{#1}}}

\def\+#1{\underline{\bf #1}}
\def\*#1{\underline{\underline{\bf #1}}}

\def\r#1{(\ref{#1})}
\def\l#1{\label{#1}}
\def\c#1{\cite{#1}}

\def\le{\left(}
\def\ri{\right)}
\def\les{\left[}
\def\ris{\right]}
\def\lec{\left\{}
\def\ric{\right\}}

\def\.{\mbox{ \tiny{$^\bullet$} }}

\def\epso{\epsilon_{\scriptscriptstyle 0}}

\def\muo{\mu_{\scriptscriptstyle 0}}

\def\tr{(ct,\#r)}
\def\ok{(\omega/c,\#k)}

\def\eps{\epsilon}

\def\ct{\cos\theta}

\def\curl{\nabla\times}

\def\curl{\nabla\times}

\begin{document}

\begin{center}

{\bf {\Large  Global and local perspectives of gravitationally assisted
negative--phase--velocity propagation of electromagnetic
waves  in vacuum
}}

 \vspace{10mm} \large
 Akhlesh  Lakhtakia\footnote{Fax: +1 814 863 4319; e--mail: akhlesh@psu.edu; also
 affiliated with Department of Physics, Imperial College, London SW7 2 BZ, UK}\\
 {\em CATMAS~---~Computational \& Theoretical
Materials Sciences Group\\ Department of Engineering Science and
Mechanics\\ Pennsylvania State University, University Park, PA
16802--6812, USA}\\
\bigskip
\vspace{10mm} \large
Tom G. Mackay\footnote{Corresponding Author. Fax: + 44 131
650 6553; e--mail: T.Mackay@ed.ac.uk.} and  \large
Sandi Setiawan\footnote{Fax: + 44 131
650 6553; e--mail: S.Setiawan@ed.ac.uk.}\\
{\em School of Mathematics,
University of Edinburgh, Edinburgh EH9 3JZ, UK}\\

\end{center}

\vspace{4mm}

\normalsize

\begin{abstract}

Consistently with the Einstein equivalence principle and using an
electromagnetic formulation first suggested by Tamm, we show that
a local observer cannot observe negative--phase--velocity (NPV)
propagation of electromagnetic waves in vacuum, whereas a global
observer can appreciate that phenomenon. Using the specific
example of the Kerr metric, we also demonstrate the possibility of
  NPV propagation  within
the ergosphere of a rotating black hole.

\end{abstract}

\vspace{4mm}


\vspace{4mm}


\vspace{4mm}

\section{Introduction}

Recently, we have shown that electromagnetic plane waves can
propagate in classical vacuum in such a way that the phase
velocity vector has a negative projection on the time--averaged
Poynting vector, provided that the vacuum is nontrivially affected
by a gravitational field \c{ML04_GTR}. Our approach is based upon
Tamm's electromagnetic formulation \c{Tamm}, involving  a
gravito--magnetic metric. The negative--phase--velocity (NPV)
propagation characteristic depends on the direction of the
propagation wavevector but not on the frequency.  NPV propagation
characteristics in certain homogeneous material mediums lead to
the phenomenon of negative refraction \c{LMW}--\c{Smi}, which
suggests the possibility of gravitationally assisted negative
refraction of electromagnetic waves by vacuum, with attendant
scientific implications \c{ML04_GTR},\c{ML04b}.

Researchers have studied electromagnetic wave propagation in terrestrial
 environments for several centuries,
and our planet is certainly affected by the solar gravitational field. Yet, NPV propagation in classical vacuum
has never been reported in the terrestrial context. Although the solar gravitational field evidently does not satisfy
the conditions for NPV propagation that were reported  elsewhere  \c{ML04_GTR}, evidence accumulated from the
Pioneer 10 mission has prompted suggestions that
new physics may be found even at the outer reaches of our solar system \c{ALLLNT}.
These and other considerations beg the question: Are global and local
perspectives of gravitationally assisted NPV propagation different?
In this communication, we answer that question in the affirmative.

\section{Maxwell equations in universal spacetime}

In the absence of charges and currents, electromagnetic fields obey the covariant
Maxwell equations\footnote{Roman indexes take the values 1, 2 and
3; Greek indexes take the values 0, 1, 2, and 3; summation is
implied over any repeated index; and Gaussian units are used.}
\begin{equation}
\label{ME1}
f_{\alpha\beta;\nu} + f_{\beta\nu;\alpha}+f_{\nu\alpha;\beta} = 0\,,\quad
h^{\alpha\beta}_{\quad;\beta} = 0\,,
\end{equation}
where
$f_{\alpha\beta}$ and $h^{\alpha\beta}$ are, respectively, the covariant and the contravariant
electromagnetic field tensors whereas the subscript $_{;\nu}$ indicates the covariant derivative with respect
to the $\nu$th spacetime coordinate. The spacetime~---~denoted by the vector $x^\alpha$ containing  the
normalized time coordinate $x^0 = ct$ (with $c$ as the maximum speed of light in the absence
of the gravitational field) and the space coordinates
$x^1$, $x^2$, and $x^3$~---~is Riemannian, with the metric  $g_{\alpha\beta}$ being a function of spacetime
and carrying the signature $(+,-,-,-)$ \c{Schutz}.

It is commonplace to follow up on a suggestion of Tamm
\c{Tamm},\c{Mashoon}--\c{Plebanski} and change the form of \r{ME1}
for application to electromagnetic fields in vacuum. The Maxwell
equations \r{ME1} may be expressed in noncovariant form as
\begin{equation}
\label{ME1_noncov} f_{\alpha\beta,\nu} +
f_{\beta\nu,\alpha}+f_{\nu\alpha,\beta} = 0\,,\quad \les \le -g
\ri^{1/2}  h^{\alpha\beta} \ris_{,\beta} = 0\,,
\end{equation}
wherein $g = \mbox{det} \les g_{\alpha \beta} \ris $ and the
subscript $_{,\nu}$ denotes ordinary differentiation with respect
to  the $\nu$th spacetime coordinate. We note that generalizing
the Maxwell equations from noncovariant to covariant formulations
is not totally unambiguous  \c{weinberg},\c{lightman}. In the
absence of experimental evidence to  eliminate this ambiguity, we
adopt the standard generalization \r{ME1}.

Let us introduce the
electromagnetic field vectors $E_\ell$, $B_\ell$, $D_\ell$ and
$H_\ell$ via the standard decompositions
\begin{equation}
\left.
\begin{array}{ll}
E_\ell = f_{\ell 0}\,, \quad & B_\ell = (1/2) \varepsilon_{\ell mn}f_{mn}\\
D_\ell= \le -g \ri^{1/2} h^{\ell 0}\,, \quad & H_\ell=(1/2)
\varepsilon_{\ell mn} \le -g \ri^{1/2} h^{mn}
\end{array}\right\}\,,
\label{bbb}
\end{equation}
 with $\varepsilon_{\ell mn}$ being the three--dimensional
Levi--Civita symbol. Thereby the noncovariant Maxwell equations
\r{ME1_noncov} assume the familiar form
\begin{equation}
\label{ME2}
\left.
\begin{array}{cc}
B_{\ell,\ell} = 0\,, & B_{\ell,0} + \varepsilon_{\ell mn} E_{m,n} = 0\\
D_{\ell,\ell} = 0\,, & -D_{\ell,0} + \varepsilon_{\ell mn} H_{m,n} = 0
\end{array}\right\}\,.
\end{equation}

In vacuum, the components of the electromagnetic field tensors are connected by the constitutive relations
\begin{equation}
\label{CR1} h^{\alpha\beta} =
g^{\alpha\mu}\,g^{\beta\nu}\,f_{\mu\nu}\,, \quad f_{\alpha\beta} =
g_{\alpha\mu}\,g_{\beta\nu}\,h^{\mu\nu}\,.
\end{equation}
 These constitutive relations of vacuum can be stated for the
electromagnetic field vectors as
\begin{equation}
\label{CR2}
\left.
\begin{array}{l}
D_\ell = \gamma_{\ell m} E_m + \eps_{\ell mn}\,\Gamma_m\,H_n\\[6pt]
B_\ell =    \gamma_{\ell m} H_m - \eps_{\ell mn}\, \Gamma_m\,  E_n
\end{array}\right\}\,,
\end{equation}
where
\begin{eqnarray}
\label{akh1} \gamma_{\ell m}
&=& - \le -{g} \ri^{1/2} \, \frac{{g}^{\ell m}}{{g}_{00}}, \\
 \Gamma_m &=& \frac{g_{0m}}{g_{00}}\,.
\end{eqnarray}

Equations \r{ME1} and \r{CR1} employ curved spacetime. So do
\r{ME2} and \r{CR2}, but the difference is that they look like the
familiar electromagnetic equations in flat spacetime applied to an
instantaneously reacting, bianisotropic medium. Techniques
commonly employed to handle electromagnetic problems in the
absence of gravitational fields should therefore be useful for
solving \r{ME2} and \r{CR2}. Before proceeding further with our
analysis, let us therefore recast \r{ME2} and \r{CR2} using the
conventional 3--vectors and 3$\times$3 dyadics as
\begin{equation}
\label{eq1} \left.
\begin{array}{l}
\curl \#E\tr + \displaystyle{\frac{\partial}{\partial t}} \#B\tr =
0\,\\ \vspace{-8pt} \\ \curl\#H\tr -
\displaystyle{\frac{\partial}{\partial t}} \#D\tr = 0\,
\end{array}
\right\}
\end{equation}
and
\begin{equation}
\left.
\begin{array}{l}
 \label{eq2}
\#D\tr = \epso\,\=\gamma\tr\cdot \#E\tr - \displaystyle{\frac{1}{c}}\, \#\Gamma\tr\times \#H\tr\,\\
\vspace{-8pt} \\ \#B\tr = \muo\,\=\gamma\tr\cdot \#H\tr +
\displaystyle{\frac{1}{c}}\,\#\Gamma\tr\times \#E\tr\,
\end{array}
\right\},
\end{equation}
wherein $\=\gamma\tr$ is the dyadic--equivalent of $\gamma_{\ell
m}$,
 $ \#\Gamma\tr$ is the vector--equivalent of $\Gamma_m$ and SI
 units are adopted.

\section{Global perspective}

Suppose that we wish to solve these equations in a certain region $\cal X$ of spacetime, subject to specific boundary conditions.
A fairly standard procedure would be to partition $\cal X$ into subregions $^{(n)}{\cal X}$, ($n=1,2,3,\dots$), in each of which
we would replace the nonuniform metric $g_{\alpha\beta}$ by the uniform metric $^{(n)}\tilde{g}_{\alpha\beta}$.
Correspondingly, the nonuniform vector with components $\Gamma_\ell$ would be replaced by the uniform vector with components
${}^{(n)}\tilde{\Gamma}_\ell$ in each subregion $^{(n)}{\cal X}$.
The
spacetime $x^\alpha$ would still remain curved.
After solving
 \r{ME2} and \r{CR2} in each subregion, we could stitch back the subregional solutions into the regional solution. This
 piecewise uniform approximation technique is very common for solving differential equations with nonhomogeneous
 coefficients  \c{Hoff}.

 Accordingly, let us examine the electromagnetic fields
 in the $n$th subregion. A three--dimensional Fourier transform of the
 electromagnetic field vectors can be taken, with the wavevector $\#k$ denoting
 the Fourier variable corresponding to $\#r$; thus,
 \begin{equation}
\left.
\begin{array}{l}
\#E (ct, \#r) = \displaystyle{ \frac{1}{c} \int_{-\infty}^\infty
\int_{-\infty}^\infty \int_{-\infty}^\infty
 \#{\sf E}(\omega/c, \#k) \exp\les i(\#k \. \#r - \omega t)\ris \, d\omega \, dk_1  \, dk_2\,} \\ \vspace{-2mm} \\
\#H (ct, \#r)  = \displaystyle{ \frac{1}{c} \int_{-\infty}^\infty
\int_{-\infty}^\infty  \int_{-\infty}^\infty
 \#{\sf H} (\omega/c, \#k) \exp\les i(\#k \. \#r - \omega t )\ris \, d \omega \, dk_1 \, dk_2\, }
\end{array}
\right\},
\label{der}
  \end{equation}
 where $i=\sqrt{-1}$ and $\omega$ is the
 usual temporal frequency. The wavevector component
  $k_3$ is determined by first substituting \r{der} in \r{eq1}, and then formulating a 4$\times$4 matrix ordinary
 differential equation which is then solved as an eigenvalue problem \c{Lopt92}.

As our focus lies here on propagating plane waves,
evanescent  (nonuniform) solutions are eliminated from further consideration by enforcing the restriction
$k_{3} \in \mathbb{R}$.
 We  remark that  $\#{\sf E} (\omega/c, \#k)$ and  $\#{\sf H} (\omega/c, \#k)$ are
  the complex--valued phasors of the electric and magnetic fields,
  respectively,
associated with a plane wave, for which a time--averaged Poynting
vector $ \langle \, \#{\sf P} (\omega/c, \#k) \, \rangle_t$ can be
derived. When the projection of $\#k$ on $\langle \, \#{\sf
P}(\omega/c, \#k) \, \rangle_t$ is negative, i.e., $\#k \. \langle
\, \#{\sf P}(\omega/c, \#k) \, \rangle_t < 0$, we say that the
phase velocity is negative.

The crucial quantity $\#k \. \langle \, \#{\sf P}(\omega/c, \#k)
\, \rangle_t $ is derived as follows  \c{ML04_GTR}:  Combining the Fourier
representations \r{der} with the constitutive relations \r{eq2}
and the Maxwell curl postulates \r{eq1}, we get
\begin{eqnarray}
\#p \times \#{\sf E}\ok &=& \omega\muo \,\,{}^{(n)}\={\tilde\gamma}
\.\#{\sf H}\ok\,, \l{h12}
\\
\#p \times \#{\sf H}\ok &=& -\omega\epso \,\,{}^{(n)}\={\tilde\gamma}
\. \#{\sf E}\ok\,\l{e12}
\\
\end{eqnarray}
where
\begin{equation}
\#p = \#k -  \frac{\omega}{c} \,\,{}^{(n)}\#{\tilde\Gamma}\,.
\end{equation}
The use of \r{h12} to eliminate $\#{\sf H}\ok$ from \r{e12} provides,
after some manipulation, the eigenvector equation
\begin{equation}
\lec \les \le\frac{\omega}{c}\ri^2 \, \vert
{}^{(n)}\={\tilde\gamma}\vert - \#p\.
{}^{(n)}\={\tilde\gamma}\cdot\#p\ris\=I +\#p\,\#p\.
{}^{(n)}\={\tilde\gamma}\ric \cdot \#{\sf E}\ok  = \#0\,
 \label{eee2}
 \end{equation}
 for representing $\#{\sf E}\ok$,
and the corresponding dispersion relation
\begin{equation}
\label{dispeq} \les\#p \. \,\,{}^{(n)}\={\tilde\gamma} \. \#p -
\le\frac{\omega}{c}\ri^2 \, \vert \,\,{}^{(n)}\={\tilde\gamma}
\vert\ris^2 = 0\,.
\end{equation}
Herein, $\=I$ is the identity
 dyadic, and $\vert{}^{(n)}\={\tilde\gamma}\vert$ is
  the determinant of
 ${}^{(n)}\={\tilde\gamma}$. Considering \r{eee2} in light of
 \r{dispeq}, we see that all $\#{\sf E}\ok$ eigenvectors solutions must satisfy the relation
\begin{equation}
\#p\.{}^{(n)}\={\tilde\gamma}\. \#{\sf E}\ok = 0\,. \label{eee3}
\end{equation}

Let us introduce the eigenvalues  ${}^{(n)}\tilde\gamma_{1,2,3}$
and corresponding eigenvectors  ${}^{(n)}\tilde{\#v}_{1,2,3}$ of
${}^{(n)}\={\tilde\gamma}$. For the purposes of illustration,
suppose we consider planewave propagation along the direction
parallel to ${}^{(n)}\tilde{\#v}_{3}$, while the vector
${}^{(n)}\tilde{\#\Gamma}$ lies at an angle $\theta$ in the plane
of the eigenvectors  ${}^{(n)}\tilde{\#v}_{1}$ and
${}^{(n)}\tilde{\#v}_{3}$, as per
\begin{equation}
\left.
\begin{array}{l}
{}^{(n)}\tilde{\Gamma}_1 = \,\,{}^{(n)}\tilde{\Gamma} \, \sin \theta  \\
{}^{(n)}\tilde{\Gamma}_2 = 0 \\
{}^{(n)}\tilde{\Gamma}_3 = \,\,{}^{(n)}\tilde{\Gamma} \, \cos \theta
\end{array}
\right\}.
\end{equation}
After exploiting the orthogonality condition \r{eee3}, the general
solution to \r{eee2} may be expressed as the sum of two
independent modes as
\begin{equation}
\l{e_gs} \#{\sf E}\ok =
 A_a\ok\,
 \#{\sf e}_a\ok\,
+
 A_b\ok\,
 \#{\sf e}_b\ok\,
\end{equation}
where
\begin{equation}
\left.
\begin{array}{l}
 \#{\sf e}_a\ok = \,\,{}^{(n)}\tilde{\#v}_{2} \\
\#{\sf e}_b\ok = \,{}^{(n)}\tilde\gamma_{3}\,\, \le \, | \#k | -
\displaystyle{\frac{\omega}{c}} \,\,{}^{(n)}\tilde{\Gamma} \cos
\theta \, \ri \,{}^{(n)}\tilde{\#v}_{1}  \,
 + \,\,{}^{(n)}\tilde\gamma_{1} \,
\displaystyle{\frac{\omega}{c}}
 \,{}^{(n)}\tilde{\Gamma}
\,\, \sin \theta \, \,{}^{(n)}\tilde{\#v}_{3} \,
\end{array}
\right\}.
\end{equation}
 The
complex--valued scalars
 $ A_{a,b}$ are unknown amplitude functions that
 can be determined from initial and boundary
conditions. The corresponding general solution for $\#{\sf H}\ok$
follows straightforwardly by combining \r{e_gs} with the Maxwell
curl postulates \r{eq1} as
\begin{equation}
\l{h_gs} \#{\sf H}\ok =
 A_a\ok\,
 \#{\sf h}_a\ok\,
+
 A_b\ok\,
 \#{\sf h}_b\ok\,
\end{equation}
with
\begin{equation}
\left.
\begin{array}{l}
 \#{\sf h}_a\ok =
\displaystyle{\frac{1}{\omega \muo}} \,\,{}^{(n)}\={\tilde\gamma}^{-1}
\. \les \le \displaystyle{\frac{\omega}{c}} \,\,{}^{(n)}\tilde{\Gamma}
\cos \theta  - | \#k | \ri \,\,{}^{(n)}\tilde{\#v}_{1} -
\displaystyle{\frac{\omega}{c}} \,\,{}^{(n)}\tilde{\Gamma}
  \,\sin \theta \,\,{}^{(n)}\tilde{\#v}_{3} \,\ris \\ \vspace{-4pt} \\
\#{\sf h}_b\ok = \displaystyle{\frac{1}{\omega \muo}}
\,\,{}^{(n)}\={\tilde\gamma}^{-1} \. \les \le | \#k | -
\displaystyle{\frac{\omega}{c}} \,\,{}^{(n)}\tilde{\Gamma} \cos \theta
 \ri^2 \,\,{}^{(n)}\tilde{\gamma}_{3} + \le \displaystyle{\frac{\omega}{c}}
\,\,{}^{(n)}\tilde{\Gamma} \sin \theta \ri^2
\,\,{}^{(n)}\tilde{\gamma}_{1} \,\ris
 \,\,{}^{(n)}\tilde{\#v}_{2}
\end{array}
\right\}.
\end{equation}

The dispersion equation \r{dispeq} reduces to a $| \#k
|$--quadratic expression which yields the two wavenumbers
\begin{equation}
\l{wn} | \#k |  = \frac{\omega}{c} \le \,\,{}^{(n)}\tilde{\Gamma} \cos
\theta \pm \sqrt{ \,\,{}^{(n)}\tilde\gamma_{1}
\,\,{}^{(n)}\tilde\gamma_{2} -
\frac{\,\,{}^{(n)}\tilde\gamma_{1}}{\,\,{}^{(n)}\tilde\gamma_{3}}
\,\,{}^{(n)}\tilde{\Gamma}^2 \sin^2 \theta } \, \ri.
\end{equation}

Finally, combining \r{e_gs}, \r{h_gs} with \r{wn}, we find that
\begin{eqnarray}
\#k \. \langle \, \#{\sf P}(\omega/c, \#k) \, \rangle_t &=&
\frac{1}{2 \muo \,\,{}^{(n)}\tilde\gamma_{3} } \Bigg[ \le
 \,\,{}^{(n)}\tilde\gamma_{1} \,\,{}^{(n)}\tilde\gamma_{2}
 -\frac{\,\,{}^{(n)}\tilde\gamma_{1}}{\,\,{}^{(n)}\tilde\gamma_{3}}
  \,\,{}^{(n)}\tilde{\Gamma}^2 \sin^2 \theta
  \ri
  \nonumber \\ &&
\pm \,\,{}^{(n)}\tilde{\Gamma} \ct \sqrt{ \,\,{}^{(n)}\tilde\gamma_{1}
\,\,{}^{(n)}\tilde\gamma_{2}
 -\frac{\,\,{}^{(n)}\tilde\gamma_{1}}{\,\,{}^{(n)}\tilde\gamma_{3}}
 \,\,{}^{(n)}\tilde{\Gamma}^2 \sin^2 \theta  }  \, \Bigg] \nonumber \\ && \times \le | A_a\ok |^2 +
\,\,{}^{(n)}\tilde\gamma_{3}\,
  \omega^2\,
\vert {}^{(n)}\={\tilde\gamma}\vert \,
 | A_b \ok |^2  \ri \,. \l{kp}
\end{eqnarray}

Let us probe the physical significance of \r{kp}. To do so, we
focus upon a specific gravitomagnetic metric, namely the Kerr
metric. We begin by noting that for a rotating black hole of geometric mass
$m_{rbh}$ and angular velocity $a_{rbh}$ in conventional units \c{Inverno}, the Kerr metric yields
\begin{equation}
\vert {}^{(n)}\tilde{\=\gamma} \, \vert = \frac{\le
{}^{(n)}\tilde{R}^4 + a_{rbh}^2  \,\,{}^{(n)}\tilde{z}^2 \ri^2}{\le
{}^{(n)}\tilde{R}^4 + a_{rbh}^2 \,\,{}^{(n)}\tilde{z}^2 - 2 m_{rbh}
\,\,{}^{(n)}\tilde{R}^3 \ri^2}
\end{equation}
and
\begin{equation}
\left.
\begin{array}{l}
\,\,{}^{(n)}\tilde\gamma_{1} = \,\,{}^{(n)}\tilde\gamma_{2} =
\displaystyle{ \frac{ \,\,{}^{(n)}\tilde{R}^4 + a_{rbh}^2
\,\,{}^{(n)}\tilde{z}^2 }{ \,\,{}^{(n)}\tilde{R}^4 + a_{rbh}^2
\,\,{}^{(n)}\tilde{z}^2 - 2 m_{rbh} \,\,{}^{(n)}\tilde{R}^3 }}
\\ \vspace{-2pt}
\,\,{}^{(n)}\tilde\gamma_{3} = 1
\end{array}
\right\},
\end{equation}
 where ${}^{(n)}\tilde{x}$, ${}^{(n)}\tilde{y}$ and ${}^{(n)}\tilde{z}$ are the values
of the three space coordinates at some representative point in
the subregion $^{(n)}{\cal X}$. The quantity ${}^{(n)}\tilde{R}$
is defined implicitly via
\begin{equation}
{}^{(n)}\tilde{R}^2 = \,\,{}^{(n)}\tilde{x}^2 +
\,\,{}^{(n)}\tilde{y}^2 + \,\,{}^{(n)}\tilde{z}^2 - a_{rbh}^2\,
\les \, 1- \le\frac{{}^{(n)}\tilde{z}}{{}^{(n)}\tilde{R}}\ri^2\ris
\,.
\end{equation}
Since both
$\vert {}^{(n)}\tilde{\=\gamma} \, \vert
> 0$ and ${}^{(n)}\tilde\gamma_{3} > 0 $, we see from \r{kp} that
NPV propagation in $^{(n)}{\cal X}$ arises provided that the inequality
\begin{equation}
\vert {}^{(n)}\tilde{\=\gamma} \, \vert -
\,\,{}^{(n)}\tilde{\#\Gamma} \. \,\,{}^{(n)}\tilde{\=\gamma} \.
\,\,{}^{(n)}\tilde{\#\Gamma} < 0 \, \l{npv_cond1}
\end{equation}
is satisfied.

It is illuminating to recast \r{npv_cond1} in terms of the metric
components as per
\begin{equation}
\frac{\,\,{}^{(n)}\tilde{g}^{00}}{\,\,{}^{(n)}\tilde{g}_{00}} < 0\,.
\l{npv_cond2}
\end{equation}
For the Kerr metric we have \c{Inverno}
\begin{equation}
\frac{{}^{(n)}\tilde{g}^{00}}{{}^{(n)}\tilde{g}_{00}} =
\frac{{}^{(n)}\tilde{R}^4 + a_{rbh}^2 \,\,{}^{(n)}\tilde{z}^2 + 2 m_{rbh}
\,\,{}^{(n)}\tilde{R}^3}{{}^{(n)}\tilde{R}^4 + a_{rbh}^2 \,\,{}^{(n)}\tilde{z}^2
- 2 m_{rbh} \,\,{}^{(n)}\tilde{R}^3}.
\end{equation}
 Within the ergosphere~---~which lies between the
outer event horizon and the stationary limit surface~---~we have
\begin{equation}
m_{rbh} + \sqrt{m_{rbh}^2 - a_{rbh}^2} < \,{}^{(n)}\tilde{R} < m_{rbh}
 + \sqrt{m_{rbh}^2 - a_{rbh}^2\,\le\frac{{}^{(n)}\tilde{z}}{{}^{(n)}\tilde{R}}\ri^2
},
\end{equation}
and $m_{rbh}^2 > a_{rbh}^2$. Therefore, inside the ergosphere
\begin{equation}
\left.
\begin{array}{l}
{}^{(n)}\tilde{R}^4 + a_{rbh}^2 \,\,{}^{(n)}\tilde{z}^2 - 2 m_{rbh}
\,\,{}^{(n)}\tilde{R}^3 < 0\, \\
{}^{(n)}\tilde{R} > 0
\end{array}
\right\},
\end{equation}
and the NPV condition \r{npv_cond2} is evidently satisfied for the chosen wave--propagation case.

Thus,
when the traversal of an electromagnetic signal is traced from
point $P \in$  $^{(p)}{\cal X}$ to point $Q \in$ $^{(q)}{\cal X}$,
with $p \neq q$, the possibility of NPV propagation in some
subregions of $\cal X$ cannot be ruled out {\em a priori}.

\section{Local perspective}

The previous two sections are from a global perspective.
Let us
now attend to the local perspective of an observer located at some point $\wp$.
This observer is constrained to
formulate an electromagnetic theory valid only in some small neighborhood of $\wp$,
and this theory must emerge from observations made only in
that neighborhood. The gravitational field in this neighborhood can be
held to be virtually uniform, the metric from the global
perspective being $g_{\alpha \beta}\vert_\wp$. However, given the
admissibility of the uniform--gravity theory for all local
observers in this neighborhood, we show here that the local
observer would end up formulating a local metric~---~which is the
same as in the special theory of relativity.

By virtue of the Einstein equivalence principle \c{Lor}, $g_{\alpha \beta}\vert_\wp$ is constrained to be
such that there exists a matrix $\Lambda^{\alpha}_{\beta} $
yielding
\begin{equation}
\eta_{\mu\nu} = \Lambda^\alpha_{\mu}\,\,
{g}_{\alpha \beta}\vert_\wp \,\,\Lambda^\beta_{\nu} \,, \l{trans_1}
\end{equation}
where $\eta_{\mu \nu} = \mbox{diag} \, \les 1, -1, -1, -1 \ris$ is
the Lorentzian spacetime metric \c{Schutz}. The matrix
$\Lambda^{\alpha}_{\beta} $ can be used to construct a
flat spacetime $x'^\alpha$ for all $x^\alpha$ in the specific neighborhood of $\wp$
as per
\begin{equation}
\left.
\begin{array}{l}
\Lambda^{\alpha}_{\beta} =  \displaystyle{ \frac{\partial x'^{\alpha}}{\partial x^\beta}\,}
\\  \\
dx^0 dx^1 dx^2 dx^3  = \mbox{det} \les \Lambda^{\alpha}_{\beta} \ris dx'^0 dx'^1 dx'^2 dx'^3
\end{array}
\right\}.
\end{equation}
We must note that, as  $\Lambda^\alpha_{\mu}$ depends upon
 ${g}_{\alpha \beta}\vert_\wp$ and therefore on $x^\alpha\vert_\wp$,
  a global coordinate transformation which
simultaneously transforms ${g}_{\alpha \beta}$  into
$\eta_{\mu\nu}$ for all $x^\alpha\in{\cal X}$ cannot be realized unless gravity is totally ignored \c{Schutz}.

In the local coordinate system $x'^{\alpha}$, the spacetime metric in the specific neighborhood is given as $\eta_{\alpha \beta}$.
Repeating the steps outlined in Section~2
 with  $ \eta_{\alpha \beta}$ substituting for
 $g_{\alpha \beta}$,
we find that
the local constitutive relations for the neighborhood
reduce to the
trivial form
\begin{equation}
\label{CR_local} \left.
\begin{array}{l}
D'_\ell =  E'_\ell\\[6pt]
B'_\ell =   H'_\ell
\end{array}\right\}\,
\end{equation}
with respect to $x'^{\alpha}$. NPV propagation is not possible for the
medium characterized by the constitutive relations \r{CR_local} \c{ML04_STR}.

\section{Concluding remarks}

Thus,
we have shown that gravitationally assisted negative--phase--velocity propagation in vacuum can be appreciated only
from a global perspective based on curved spacetime; whereas a local observer, constrained to a flat
spacetime, must conclude NPV propagation is impossible in vacuum. The diversity of the local and the global perspectives
is in full accord with the Einstein equivalence principle, and can be
explained as follows:
The subregional metric ${}^{(m)}\tilde{g}_{\alpha\beta}$ appears from a suitable subregional averaging of $g_{\alpha\beta}$.
Whereas $g_{\alpha\beta}$ must satisfy the dictates of the Einstein equivalence principle at every $x^\alpha\in{\cal X}$, there is no reason for
${}^{(m)}\tilde{g}_{\alpha\beta}$ to satisfy the same dictates at even one $x^\alpha\in\,{}^{(m)}{\cal X}$.

The feasibility of NPV propagation within the ergosphere of a
rotating black hole has been demonstrated in this communication. A detailed numerical
study is called for in order to further explore the circumstances
 and implications of such type of propagation, which we hope to report in due course of time.

\medskip

\noindent{\bf Acknowledgements.} We gratefully acknowledge
discussions with Dr. Martin W. McCall of Imperial College, London.
We thank an anonymous referee for helpful suggestions, especially
relating to the Kerr metric. SS acknowledges EPSRC for support
under grant GR/S60631/01.


\begin{thebibliography}{99}

\bibitem{ML04_GTR}
A. Lakhtakia, T.G. Mackay,
 J. Phys. A: Math. Gen.  37 (2004) L505; corrigendum 37 (2004)
 12093.


\bibitem{Tamm} I.E. Tamm, Zhurnal Russkogo Fiziko-Khimicheskogo Obshchestva, Otdel Fizicheskii (J.
Russ. Phys.-Chem. Soc, Phys. Section)
 56 (1924) 248.

\bibitem{LMW}
A. Lakhtakia, M.W. McCall, W.S. Weiglhofer, in:  W.S. Weiglhofer, A. Lakhtakia (Eds.),
 Introduction to Complex Mediums for Optics and
Electromagnetics, SPIE Press, Bellingham, WA, USA, 2003, pp. 347--363.

\bibitem{PS04}
J.B. Pendry, D.R. Smith,
Phys. Today 57 (2004) 37 (June issue).


\bibitem{Smi}
D.R. Smith, Phys. World 17 (2004) 23 (May issue).

\bibitem{ML04b}
T.G. Mackay, A. Lakhtakia,
Curr. Sci.  86 (2004) 1593.

\bibitem{ALLLNT}
J.D. Anderson, P.A. Laing, E.L. Lau, A.S. Liu, M.M. Nieto, S.G. Turyshev,
Phys. Rev. D 65 (2004) 082004.

\bibitem{Schutz}
B.F. Schutz, A First Course in General Relativity, Cambridge Univ. Press, Cambridge, UK, 1985, chap. 6.



\bibitem{Mashoon}
D. Bini, C. Cherubini, B. Mashhoon, Class. Quantum Grav. 21 (2004) 3893.

\bibitem{Skrotskii}
G.V. Skrotskii, Soviet Phys.--Dokl. 2 (1957)  226.


\bibitem{Plebanski}
J. Plebanski, Phys. Rev. 118 (1960) 1396.

\bibitem{weinberg}
S. Weinberg, Gravitation and Cosmology,
John Wiley \& Sons, New York , USA, 1972, chap. 7.

\bibitem{lightman}
A.P. Lightman, W.H. Press, R.H. Price and S.A.
Teukolsky, Problem Book in Relativity and Gravitation,
 Princeton University Press, Princeton, USA, 1975, chap. 14.

\bibitem{Hoff}
J.D. Hoffman, Numerical Methods for Engineers and
Scientists, McGraw--Hill, New York, USA, 1992.

\bibitem{Lopt92}
A. Lakhtakia, Optik 90 (1992) 184.


\bibitem{Inverno}
R. d'Inverno, Introducing Einstein's Relativity, Claredon Press,
Oxford, UK, 1992, chap. 19.

\bibitem{Lor}
H.A. Lorentz {\em et al.\/}, The Principle of Relativity, Dover Publications,
New York, USA, 1952,  p.120.


\bibitem{ML04_STR}
T.G. Mackay, A. Lakhtakia,
J. Phys. A: Math. Gen.  37 (2004) 5697.


\end{thebibliography}
\end{document}